\documentstyle[12pt,epsf,epsfig,here
]{article}

\begin{document}

\title{
{\bf Uncertainty of the two-loop RG upper bound on the Higgs
mass}\thanks{IHEP 2002--21}}
\author{Yu.\ F.\ Pirogov\ and\ O.\ V.\ Zenin\\[0.5ex]
{\it Institute for High Energy Physics, Protvino, 
Moscow Region, Russia}\\
}
\date{}
\maketitle
\abstract{
\noindent
A modified criterion of the SM perturbative consistency 
is proposed. It is based  on the analytic properties of the
two-loop SM running couplings. Under the criterion  adopted, the Higgs
mass up to 380~GeV might not give rise to the strong coupling prior to
the Planck scale.
This means that the light Higgs boson is possibly preferred for
reasons other than the SM perturbative consistency, 
i.e.\ for reasons beyond the SM.
}

\subsection*{1\quad Introduction}

The current experimental data restrict the Higgs mass in the 
Standard Model (SM) within the range  $114.1$~GeV$ < M_H < 194$~GeV.
The lower bound on $M_H$  comes from the absence of the Higgs
production signal at LEP II at the 95\% CL~\cite{LEP}.
The upper bound is derived at the same CL from the fit to the
precision electroweak data~\cite{Precision}.
On the other hand, the upper bound on the Higgs mass can be 
obtained from the requirement of the SM perturbative consistency
up to a cutoff energy scale $\Lambda$ at
which the SM might get into the strong coupling regime. 
The two-loop renormalization group (RG) 
gives typical upper bounds $M_H < 200$~GeV at 
$\Lambda = M_{GUT} = 10^{14}$~GeV and
$M_H < 180$~GeV at $\Lambda = M_{Pl} = 10^{19}$~GeV
(see, e.g., \cite{Pirogov98}). 
Thus both the electroweak precision data and 
the SM perturbative consistency up to the GUT scale 
exclude the Higgs mass $M_H \ge 200$~GeV.
This could  be interpreted as though the Higgs should be  
light due to the self-suppression of the strong coupling in the SM.
But the question is to what extent the Higgs
upper bound  from the SM perturbative consistency
is reliable?

A clear-cut criterion of the strong coupling in the Higgs sector of
the SM exists only in one loop. In this case, the one-loop 
quartic coupling $\lambda$ develops the Landau pole at a finite energy
scale $\Lambda$. In two loops, the pole is compensated but $\lambda$
becomes large, $\lambda/4\pi^2 \simeq 1$, nearly at the same energy
scale $\Lambda$. Taken alone, this does not give the unambiguous
criterion of the nonperturbative regime any more.
In the conventional 
assumption that the higher loops become comparable with the first
and second ones at the same scale $\Lambda$,
the results of~\cite{Pirogov98}, \cite{RG} follow (see also
\cite{Sher} for a review). On the other hand
the contributions of the higher loops might be either small, 
or large but mutually compensated. 
This would not change drastically the two-loop running of $\lambda$
and may relax the conventional upper bound on the Higgs mass.

Presently, the full set of the SM $\beta$ functions is known up to 
the two loops only. 
This forces one  to study the reliability
of the self-consistency criterion
of the two-loop RG approximation in the SM.
This is the purpose of the present paper.
The method proposed in the paper 
relies on the subtracted RG and 
the analytic properties of the running couplings.
It is similar in spirit to methods applied to resolve the Landau
singularity problem in QED \cite{QED} and, later,
to improve the infrared behaviour of the QCD running coupling
$\alpha_S(\mu^2)$ \cite{aleks}, \cite{QCD}.

\subsection*{2\quad Subtracted finite-loop RG}

Let us consider the system of the SM two-loop RG equations (RGE): 
\begin{equation}\label{RGE}
\mu^2 \frac{da_i(\mu^2)}{d\mu^2} = \beta_i
\left(\left \lbrace a_j(\mu^2) \right \rbrace \right)~.
\end{equation}
Here and in what follows $a_i(\mu^2)$ are the SM running  
couplings vs.\ the energy squared scale~$\mu^2$,
and $\beta_i$ are the respective $\beta$ functions calculated at the 
given number of loops. We disregard the mass effects here.
Conventionally, the system (\ref{RGE}) is integrated numerically along
the real axis $\mbox{Re}~\mu^2 < 0$:
\begin{equation}\label{a(mu)}
a_i(\mu^2) = a_i (\mu^2_0) + 
\int_{\mu^2_0}^{\mu^2} \frac{d{\mu'}^2}{{\mu'}^2} 
\beta_i \left(\left \lbrace a_j({\mu'}^2) \right \rbrace \right)~,
\end{equation}
where $\mu^2_0 < 0$ is a reference point, $\vert \mu_0\vert  \sim
M_Z$.
The $\beta$ functions can now be defined as the functions of the
real negative $\mu^2$:
\begin{equation}\label{beta(mu)}
\beta_i(\mu^2) \equiv
\beta_i \left( \left\lbrace a_j(\mu^2) \right\rbrace \right)~.
\end{equation}

Eqs.~(\ref{RGE}) -- (\ref{beta(mu)}) preserve their 
meaning for the complex $\mu^2$ as well.
But the numerical solution obtained says nothing about the analytic
properties of the running couplings with respect to $\mu^2$. 
In two loops, despite the absence of the real
singularities of the Higgs quartic coupling $\lambda$
there could be the complex ones.
They influence 
the strong coupling regime $\lambda/4\pi^2 \ge 1$ at large enough real
$\mu^2$. The extension of the two-loop RG analysis onto the complex
$\mu^2$ plane allows one to find the position of the singularities
implicitly.

To this end, let us continue analytically the $\beta$ functions and
running couplings onto the complex $\mu^2$ plane with the cut along
the real axis $\mbox{Re}~\mu^2 > 0$~(Fig.~1).
The cut is chosen so that $-\pi < \mbox{Im}~\ln{}(-\mu^2) < \pi$.
All the running couplings are assumed to satisfy the hermiticity
condition $a_i(\mu^{2*}) = a^*_i(\mu^2)$. 
Let us first choose the closed contour 
$C = C_0 \cup C_+ \cup \tilde{C} \cup C_+^*$ (Fig.~1)
so that $C$ encircles the given point $\mu^2$,
and all the  singularities of the running couplings $a_i(\mu^2)$
reside outside~$C$. 
Then $\beta_i(\mu^2)$ satisfy the identity
\begin{equation}\label{koshi}
\beta_i(\mu^2) \equiv 
\frac{1}{2\pi i} \int_C \frac{\beta_i(s)~ds}{s - \mu^2}~,
\end{equation}
where $\beta_i(s) \equiv \beta_i(\lbrace a_j(s)\rbrace{})$.
Substituting Eq.\ (\ref{koshi}) into Eq.\ (\ref{a(mu)}) one gets
\begin{equation}\label{RGE_koshi}
a_i(\mu^2) =  a_i(\mu^2_0)  + 
\frac{1}{2\pi i} \int_{\mu_0^2}^{\mu^2} \frac{d{\mu'}^2}{{\mu'}^2}
\int_C \frac{\beta_i(s)~ds}{s - {\mu'}^2}~,
\end{equation}
where the integration path between points $\mu^2_0$ and $\mu^2$ should
lie inside $C$. In what follows, the square root $\tilde{\Lambda}$ 
of the  radius of the outer contour $\tilde C$ 
is referred to as the modification radius.

\begin{figure}[H]
\vspace*{31ex}
\hspace*{-15ex}
{\epsfxsize=50mm \epsfbox[-70 40 70 160]{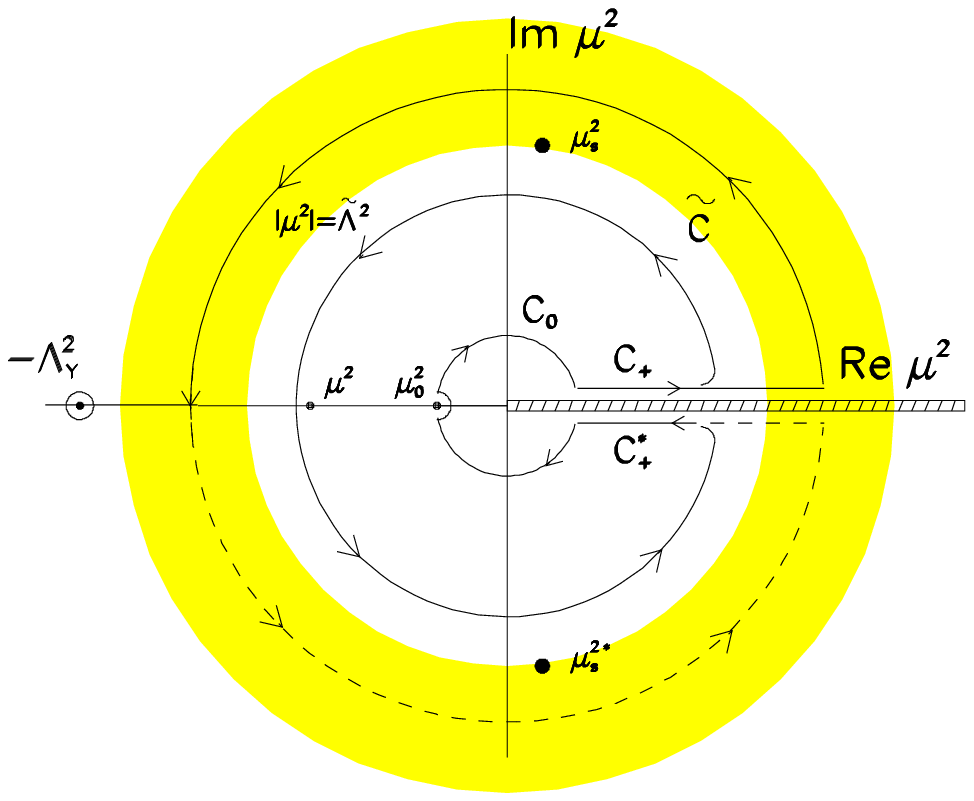}} 

\vspace*{-9ex}
{\bf Fig.~1:}~The integration contour $C$ and
the generic complex-conjugate singularity points $\mu^2_s$, 
${\mu_s^2}^*$ with $\vert \mu^2_s\vert  = \Lambda_s^2$. The real point
$(-\Lambda^2_Y)$ corresponds to the $U(1)_Y$ singularity.
All the complex singularities are assumed to reside within the 
shadowed area at $\Lambda_s^2\le\vert\mu^2\vert \le
\Lambda^2_Y$. The hatched line designates the physical cut. 
\end{figure}

Now let us spread the outer contour  $\tilde{C}$
so that at least a part of the implicit singularities of $a_i(\mu^2)$ 
gets located inside $C$. In general, 
the identity (\ref{RGE_koshi}) ceases  to be valid. 
Moreover, the integration of the RGE system (\ref{RGE})
from the reference point $\mu^2_0$ 
to the real point $(-\tilde{\Lambda}^2)$, $\tilde\Lambda>\Lambda_s$,
along the real axis
and the upper half of the contour $C$ (shown in solid in Fig.~1)
does not give the identical results.  Remarkably, in the latter
case the couplings $a_i(-\tilde{\Lambda}^2)$ acquire  the
nonzero complex parts while in the former case they are real by
construction. This discrepancy reflects the contribution  of the
implicit complex singularities. The minimal  radius $\Lambda^2_s$
of the external contour $\tilde C$ at which all these irregularities
take place gives the estimate 
of the  upper range of the reliability of the RG in the given loops. 
The value of $\Lambda_s$ corresponds to crossing the nearest
singularities of 
$a_i(\mu^2)$. At scales larger than $\Lambda_s$,
the original finite-loop approximation is definitely unreliable.
It is at $\vert \mu^2\vert  \ge \Lambda^2_s$, where the
contributions of
higher loops are needed to improve the analytic properties of
the conventional running couplings $a_i(\mu^2)$. 

The above  procedure suffices to give the clear-cut numerical 
criterion of the self-consistency of the finite-loop~RG.
But to visualise, let us  modify Eq.~(\ref{RGE_koshi}) and  define the
new running couplings 
$a^{(\tilde{\Lambda})}_i(\mu^2)$ as follows
\begin{equation}\label{a_i_reg_L}
a^{(\tilde{\Lambda})}_i(\mu^2) = 
a_i(\mu^2_0)  + 
\int_{\mu_0^2}^{\mu^2} \frac{d{\mu'}^2}{{\mu'}^2} 
\beta^{(\tilde{\Lambda})}_i({\mu'}^2)~,
\end{equation}
with the once subtracted $\beta$ functions
\begin{equation}\label{beta_subs_L}
\beta^{(\tilde{\Lambda})}_i(\mu^2) \equiv  
\beta_i(\mu^2_0) + \frac{1}{2\pi i} \int_C ds~\beta_i(s) 
\left( \frac{1}{s - {\mu}^2} - \frac{1}{s - {\mu_0}^2} \right)~.
\end{equation}
Here the point $\mu^2_0$ is shifted  infinitesimally inside $C$ and
$\beta_i(s)$, restricted to contour $C$,  
are obtained by integrating the RGE system
(\ref{RGE}) along the contour $C$ itself.
By the very construction, the modified couplings
${a}^{(\tilde\Lambda)}_i(\mu^2)$ exactly coincide
with $a_i(\mu^2)$ at $\vert \mu \vert < \tilde\Lambda$ if the
integration contour does not encompass 
the complex singularities, i.e.\  $\tilde \Lambda<\Lambda_s$.
Due to hermiticity, the couplings are real at the real
negative $\mu^2$.  
If the complex singularities get inside the contour,
the procedure is not uniquely defined. In particular
$a^{(\tilde{\Lambda})}_i(\mu^2)$ cease generally to be hermitian. To
improve this,  we redefine the integral in
Eq.~(\ref{beta_subs_L}) as the contribution of  the upper half of the
contour $C$  minus the contribution of  the symmetric lower
half of the contour calculated in the similar manner. This does
not change the results at
$\tilde\Lambda<\Lambda_s$. So defined 
$a^{(\tilde{\Lambda})}_i(\mu^2)$  are regular and hermitian and 
differ from $a_i(\mu^2)$ by the contribution
of singularities and normalization constants.
The constants are chosen so that 
$\beta^{(\tilde{\Lambda})}_i(\mu^2_0) \equiv \beta_i(\mu^2_0)$
and hence ${a}^{(\tilde\Lambda)}_i(\mu^2)= a_i(\mu^2)+ 
{\cal O}((\mu^2 - \mu_0^2)^2)$ in a~vicinity of  $\mu^2_0$ where the
finite-loop RG is believed to be reliable.  
The large difference between the couplings arises as soon as the
singular parts of $a_i(\mu^2)$ become large.

\subsection*{3\quad  Modification of the SM two-loop couplings}

The SM ultraviolet behaviour has been extensively studied by the
conventional RG method up to the two 
loops~\cite{Pirogov98} -- \cite{Sher}. 
An important outcome of this study is the range of the Higgs mass
for which the SM remains perturbatively consistent up to the given
cutoff scale $\Lambda$.
The consistency can be broken
either by the heavy enough Higgs, whose quartic
coupling $\lambda$ ``blows up'' at the scale $\Lambda$,
or by the light Higgs, whose coupling $\lambda$ dumps
below zero at the  scale $\Lambda$.\footnote{The upper
and the lower bounds on the Higgs mass are also known in the
literature as the triviality bound and the vacuum stability bound,
respectively.} 
Thus, quite a narrow corridor is retained for the Higgs mass
(see, e.g., Fig.~4 of Ref.~\cite{Pirogov98}).
These bounds are of special interest because the Higgs mass
remains the last undetermined SM parameter.

In two loops, the Higgs quartic coupling
$\lambda$, as well as the other SM couplings,
develops no  singularities prior to
the Landau singularity of the $U(1)_Y$ gauge coupling at $\Lambda
\ge 0.2 \cdot 10^{41}$~GeV, the latter corresponding to the Higgs mass
$M_H\ge 114.1$~GeV~\cite{Pirogov98}.
The situation is obscured by the fact that the SM two-loop RG
equations can be solved only numerically.
The numerical solution vs.\ real $\mu^2$
provides no information about the analytic properties
of the SM two-loop running couplings.

The method of analytic modification studies
the evolution of the running couplings vs.\ complex $\mu^2$.
The variation of the modification radius $\tilde{\Lambda}$ (Fig.~1)
allows one to determine the two-loop singularity scale $\Lambda_s$
without finding the unphysical singularities explicitly.
Thus one can judge about the self-consistency
of the two-loop RG at the given energy scale $\mu$.
It is sufficient to calculate the modified couplings
$a^{(\tilde{\Lambda})}_i(\mu^2)$ 
and compare them to the conventional ones Eq.~(\ref{a(mu)}).
This enables one to determine the radius $\Lambda_s$
at which the singularity is located,
making the numerical analysis rather productive.
If $\tilde{\Lambda}<\Lambda_s$, then the 
conventional and the modified SM running couplings 
are identical within the routine accuracy,
$a_i(\mu^2) \equiv a_i^{(\tilde{\Lambda})}(\mu^2)$,
$\vert \mu^2\vert ^{1/2}<\tilde{\Lambda}$.
As soon as $\tilde{\Lambda}$ exceeds $\Lambda_s$, 
the modified couplings depart from the respective 
conventional ones.

To illustrate, consider the two-loop RG evolution of the SM
with the maximally heavy Higgs, $M_H = 200$~GeV, 
nearly allowed by the electroweak precision data~\cite{Precision}.
Varying the modification radius $\tilde{\Lambda}$ in the range
$10^{19}~\mbox{GeV}<\tilde{\Lambda}<10^{42}~\mbox{GeV}$,\footnote{
I.e.\ well below the Landau singularity of the $U(1)_Y$ gauge coupling
at $\Lambda_s \simeq 5\cdot 10^{50}$~GeV for this $M_H$.} we
find numerically  the scale of the two-loop hidden singularity to be
$\Lambda_s \simeq 10^{31}~\mbox{GeV}$.
This can be seen from Fig.~2 showing the conventional (RG) and
subtracted (SRG)  two-loop running of the Higgs quartic coupling
$\lambda$. Note that $\lambda$ gets in fact rather large decrement, of
10\% or so, after the integration contour crosses over
the implicit singularities.
For the lighter Higgs (not shown), $\lambda$ stays 
actually unmodified.\footnote{For the 380~GeV Higgs,
the modification of the $t$, $b$, and $\tau$ Yukawa couplings (not
shown) cancels the unification of the latter ones \cite{Pirogov98}
above the singularity scale $\Lambda_s$.}
Fig.~3 shows  the conventional and modified two-loop evolution    
of the SM gauge couplings. In these figures, the modification radius
is $\tilde{\Lambda} = 10^{42}$~GeV and  $\vert \mu_0\vert$ is equal to
the Higgs VEV, $v=246.2$~GeV.
The   extension of  $\tilde{\Lambda}$ even beyond the position of the
Landau singularity  results in the relative variation of the
modified running couplings at the level of~$10^{-3}$.  
The case $M_H = 380$~GeV corresponds to 
$\Lambda_s = M_{Pl} = 10^{19}$~GeV.
Also shown in  Fig.~3 is the evolution of $\alpha_1(\mu^2)$ 
at $M_H \simeq 1.2$~TeV
which  corresponds to $\Lambda_s = M_{GUT} = 10^{14}$~GeV. 

\begin{figure}[H]
{\epsfysize=90mm \epsfbox[25 10 755 435]{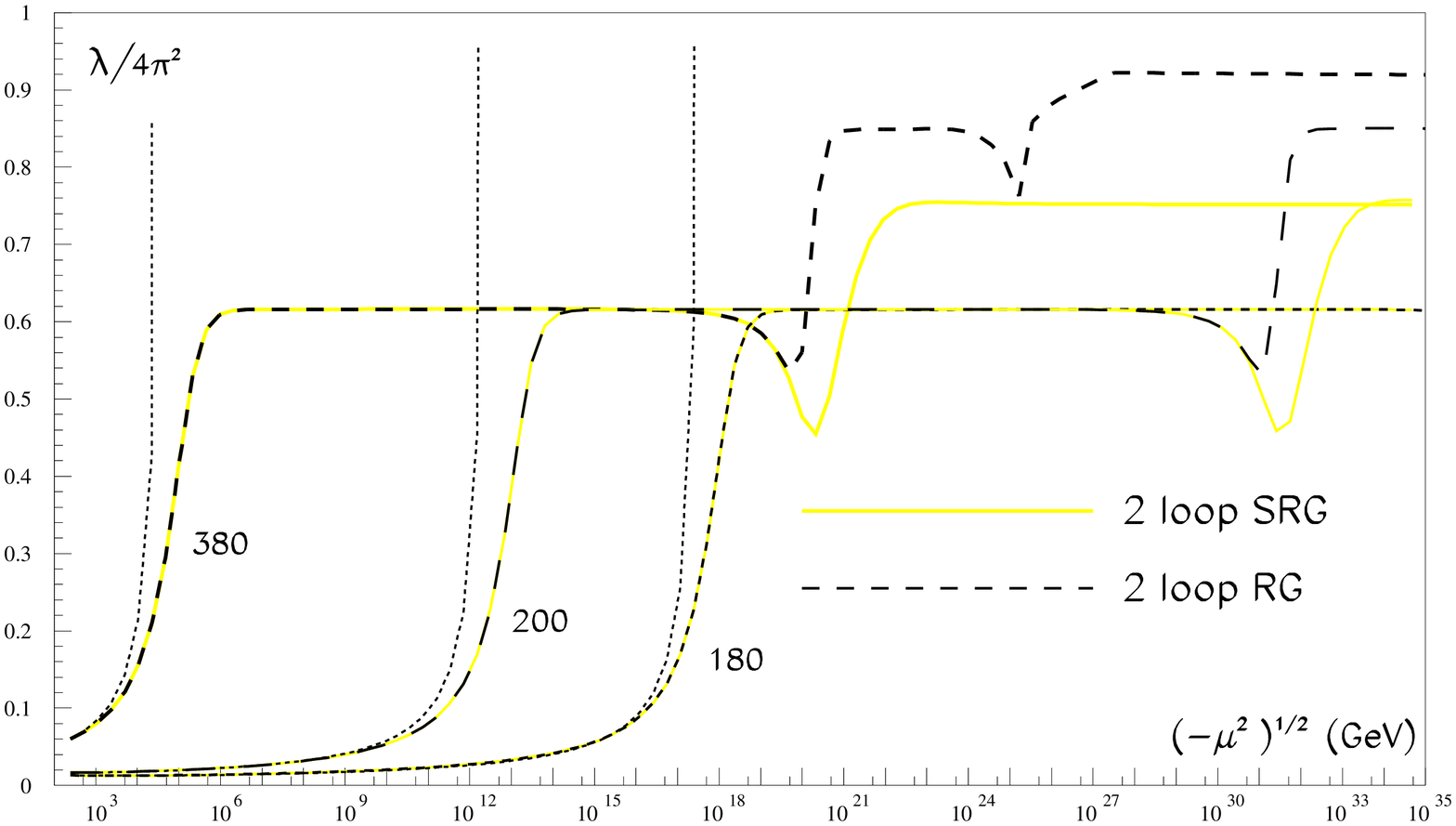}} 

{{\bf Fig.~2:} The conventional (RG) and  
subtracted (SRG) two-loop running
of the SM Higgs quartic coupling $\lambda$ at $M_H$ =180~GeV --
380~GeV. For comparison, the one-loop RG running of 
$\lambda$ is shown by dots.
}
\end{figure}

\paragraph*{The upper Higgs bound}

An important conclusion follows hereof.
For the 200~GeV Higgs, all the SM couplings demonstrate very close
conventional and modified two-loop running up to the two-loop
singularity scale $\Lambda_s$. The Higgs mass $M_H=200$ GeV
spoils the analytic properties of the SM two-loop running
couplings only at the scale $\Lambda_s \simeq 10^{31}$ GeV,
i.e.\ well above the Planck scale.
This can imply that to improve the analytic properties of
the SM two-loop couplings, 
the contributions of the third and higher loops
are needed only at scales $\mu > M_{Pl}$. To break down the
perturbativity of the SM prior to the Planck  scale
$M_{Pl}=10^{19}$~GeV the Higgs mass $M_H>380$~GeV is required.
This lifts up the commonly accepted upper bound on the Higgs mass
$M_H \le 180$~GeV derived in the conventional manner 
from the same requirement. Moreover, to guarantee the SM
perturbativity up to the GUT scale, $M_{GUT} = 10^{14}$~GeV, 
it is not actually necessary to impose any upper bound on~$M_H$.
Thus the Higgs is light probably for reasons other than 
the absence of the strong coupling in the SM. 
These reasons might lie beyond the SM.
E.g., the Higgs could be the composite
pseudo-Goldstone boson having the natural 
mass $\sim{}M_Z$ \cite{Pirogov92}.

\begin{figure}[H]
{\epsfysize=110mm \epsfbox[0 0 735 567]{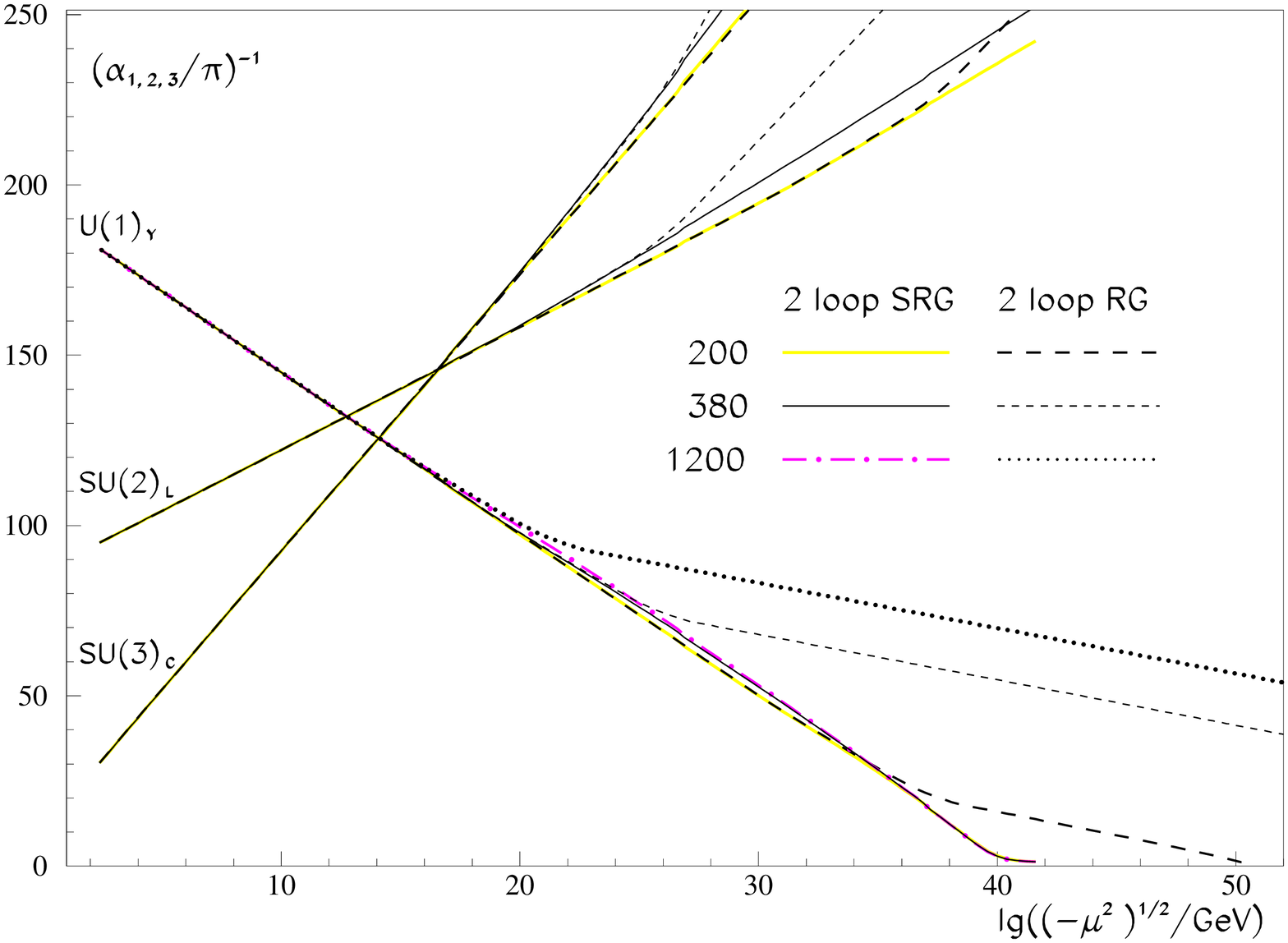}} 

{{\bf Fig.~3:} The conventional (RG) and 
subtracted (SRG) two-loop running
of the SM gauge couplings at $M_H$ = 200~GeV and 380~GeV.
$M_H$ = 380~GeV corresponds to $\Lambda_s = M_{Pl}$. 
The running of the $U(1)_Y$ gauge coupling at $M_H \simeq 1.2$~TeV
corresponding to $\Lambda_s = M_{GUT}$
is also shown.
}
\end{figure}

To resolve the uncertainty of the Higgs upper bound
the third and fourth  loops
in the SM are urgently needed.
Two extreme possibilities can be envisaged. 
First, the higher loops are large and do not compensate each other.
In this case, the conservative conventional upper bound $M_H <
180$~GeV at $\Lambda_s = M_{Pl}$ would follow.
Second, the higher loops are either small,
or large but mutually compensated.
In this case,  the more liberal modified upper
bound is appropriate,  and
$M_H$ up to 380~GeV would be  allowed at the same $\Lambda_s$.
More realistically, an intermediate case may realize so that the upper
bound on $M_H$ should lie somewhere in between 180~GeV and 380~GeV.

\paragraph*{The lower Higgs bound}

The low Higgs masses, $M_H \le 138.1$~GeV,\footnote{This
corresponds to the recalculated result
of Ref.~\cite{Pirogov98} for the central value 174.3~GeV
\cite{Hagiwara} of the top mass.}
give rise to the electroweak vacuum instability prior to the Planck
scale. However at the vacuum instability
scale, the SM running couplings develop no singularities
and hence require no subtractions.
Thus the analytic modification method taken as it is cannot
clarify the electroweak vacuum instability problem.

\subsection*{4\quad  Conclusion}

The subtracted RG is applied to study the two-loop 
self-consistency of the SM.
It is found that at the Higgs mass $M_H < 380$~GeV,
the two-loop singularity scale is 
$\Lambda_s > M_{Pl}$.
This implies that $M_H<380$~GeV does not necessarily 
threaten with the strong coupling prior to the Planck scale.
Even allowing $\Lambda_s$ as low as $M_{GUT}$,  
the SM self-consistency may actually impose 
no upper bound on $M_H$. In other words, 
the light Higgs might be preferred  for 
reasons other than the SM perturbativity, i.e.\ 
for reasons beyond the SM.
To clarify the issue the third and fourth loops in the SM RG are
needed. On the other hand the method cannot 
resolve the SM vacuum instability problem arising, in two loops,
at $M_H < 138.1$~GeV. Thus, out of the entire experimentally allowed
range for the Higgs mass 
$114.1~\mbox{GeV} < M_H < 194~\mbox{GeV}$,
only the lowest Higgs masses $114.1~\mbox{GeV}<M_H<138.1~\mbox{GeV}$
could definitely give rise to the SM inconsistency prior to the Planck
scale and would require new physics.
\vspace{-1ex}
\paragraph{Acknowledgements} The authors are grateful to 
A.~I.\ Alekseev and V.V.~Kabachenko for useful discussions.

\end{document}